\documentclass[journal]{IEEEtran}
\usepackage{graphicx}
\usepackage{amsmath}
\usepackage{txfonts}
\usepackage{bm}
\usepackage{color}
\usepackage{bbm}

\newcommand{\rank}{\mathrm{ rank } \,}
\newcommand{\e}[1]{\times 10^{#1}}

\ifCLASSINFOpdf
\else
\fi
\begin{document}
%
\title{Quantification of Propagation Modes in an Astronomical Instrument from its Radiation Pattern}
%
%
%

\author{Yasumasa~Yamasaki and~Hiroaki~Imada
\thanks{Yasumasa Yamasaki is with the Osaka Metropolitan University, Sakai 599-8531, Japan, and also with the National Astronomical Observatory of Japan, Mitaka 181-8588, Japan, e-mail: szc04116@st.osakafu-u.ac.jp, Hiroaki Imada is with the National Astronomical Observatory of Japan, Mitaka 181-8588, Japan, e-mail: hiroaki.imada@nao.ac.jp}}

%
%

\markboth{Journal of \LaTeX\ Class Files,~Vol.~14, No.~8, August~2015}%
{Shell \MakeLowercase{\textit{et al.}}: Bare Demo of IEEEtran.cls for IEEE Journals}
%



\maketitle

\begin{abstract}
In modern radio astronomy, one of the key technologies is to widen the frequency coverage of an instrument.
The effects of higher-order modes on an instrument associated with wider bandwidths have been reported, which may degrade observation precision. It is important to quantify the higher-order propagation modes, though their power is too small to measure directly. Instead of the direct measurement of modes, we make an attempt to deduce them based on measurable radiation patterns.
Assuming a linear system, whose radiated field is determined as a superposition of the mode coefficients in an instrument, 
we obtain a coefficient matrix connecting the modes and the radiated field and calculate the pseudo-inverse matrix.
To investigate the accuracy of the proposed method, we demonstrate two cases with numerical simulations, axially-corrugated horn case and offset Cassegrain antenna case, and the effect of random errors on the precision.
Both cases showed the deduced mode coefficients with a precision of $\bm{10^{-6}}$ with respect to the maximum mode amplitude and $\bm{10^{-3}}$ degrees in phase, respectively. The calculation errors were observed when the random errors were smaller than 0.01 percent of the maximum radiated field amplitude, which was a much lower level compared with measurement precision.
The demonstrated method works independently of the details of a system. The method can quantify the propagation modes inside an instrument and will be applied to most of linear components and antennas, which leads to various applications such as diagnosis of feed alignment and higher-performance feed design.
\end{abstract}


%
\IEEEpeerreviewmaketitle

\section{Introduction}
%
%
%
%
\IEEEPARstart{N}{ew} astronomical science cases are always driving the development of instruments, and technological advances in observation enable us to dig out new astronomical science cases as well.
Take several examples; \cite{Masui2021} developed a new heterodyne receiver capable of the simultaneous observation at 230 and 345\,GHz, where the CO rotational transition lines are observed.
The Atacama Large Millimeter/submillimeter Array (ALMA) 2030 development \cite{carpenter2020alma} requests
the ALMA receiver capabilities of simultaneously observing the broader radio frequency (RF) and intermediate frequency (IF) bandwidths than ever \cite{10.1117/12.2562272, Yagoubov2020, Huang2022}.
Next generation interferometers such as Square Kilometre Array (SKA) and next-generation Very Large Array \cite{Pellegrini2021, 10.1117/12.2312089} are planned in a wider frequency band, from $ \sim 100 $\,MHz up to $ \sim 100 $\,GHz.
Precise observation of polarization state is also getting more and more significant to various communities in the coming decades \cite{Hagiwara2022,LiteBIRD,CMB-S4}.
One of the key words common to those science cases is a wider bandwidth than ever to improve observation efficiency and reduce systematics.
In addition to a bandwidth, some of them also require the capability of observing polarization states.

Recent remarkable RF technological advances enable us to build an instrument with broader observation bandwidths according to the science demands.
This is because we can handle complex structure in an optimization process and access manufacturing techniques for such structure in the last decade \cite{7481872, 2021JIMTW..42..960G}.
Electromagnetic (EM) simulations have greatly supported optimizations and sometimes help us reveal the difference of measured data from the projected performance.
The literature \cite{Villiers2021, Villiers2022, Villiers2023} reported the fully-polarized beam patterns of MeerKAT, which is a precursor of SKA-Mid, using holography, and found the unidentified patterns that their EM simulations did not predict.
Therefore, we need a method to observe what happens in a component and to assess those measured data in detail.

A feed horn is one of the key components to achieve high aperture efficiency and low noise temperature over a broad frequency band.
In many cases, EM simulations of a feed and an RF system pump a fundamental mode into a source port at the edge of a transmission line.
There are some situations where we are annoyed by higher-order modes in a waveguide at higher frequencies.
One is the operation of a waveguide-type feed with a fractional RF bandwidth of $ > 60\% $.
Another case is a feed horn followed by another waveguide component such as orthomode transducer (OMT).
The latter case will see higher-order modes excited between the OMT and horn.
Higher-order modes in a waveguide give rise to beam distortions, beam squints (a few percent of their beam widths), and cross-polarization asymmetries.
Since these effects are critical for astronomical observations, it is essential to understand the EM fields at the input of the feed.
However, it is difficult to measure it directly.

Even if we limit ourselves to the fundamental modes, circular and square waveguides, respectively, have a pair of fundamental modes that are orthogonal to each other.
When both modes are excited due to misalignment, deformation, and suchlike, we will most likely struggle to improve a so-called cross-polarization property.
\cite{Carozzi2011} introduced a fundamental figure of merit, intrinsic cross-polarization ratio, to characterize a radio telescope in terms of polarimetric performance.
Since the intrinsic cross-polarization ratio is based on the Jones matrix, it perfectly describes the system performance regarding polarization but does not aim at diagnoses of RF components.
Examples of antenna system diagnostic, \cite{Borries2021, Ericsson2022, Jørgensen2010}, reported the method to reconstruct a source current based on the inverse method of moments (MoM) technique.
They introduced an imaginary closed surface encompassing the whole system and reconstructed equivalent currents on that surface from the outside radiation.
Their method would be suitable for a complex and large system such as a satellite body rather than a small component.
To handle polarization and modes at the component level, we would need a more flexible method to deduce and quantify the excited EM field of the feed from the measurable radiated field.

This work addresses the deduction of a source EM field from a radiated field. 
If a source field can be fully described with modes in a waveguide, for example, we also aim to draw out the ratios between the source modes.
Section~\ref{sec:principle} arranges assumptions and formalism for our source EM field deduction.
Section~\ref{sec:demonstration} demonstrates the developed method together with numerical simulations and addresses the effects of errors in the input radiated field on the deduction.
Some applications are discussed in Sect.~\ref{sec:discussion}, followed by Conclusions.

\section{Principle of mode estimation}\label{sec:principle}
\subsection{Assumptions} \label{sec:assumptions}
\begin{figure}[t]
 \begin{center}
 \includegraphics[width=8cm]{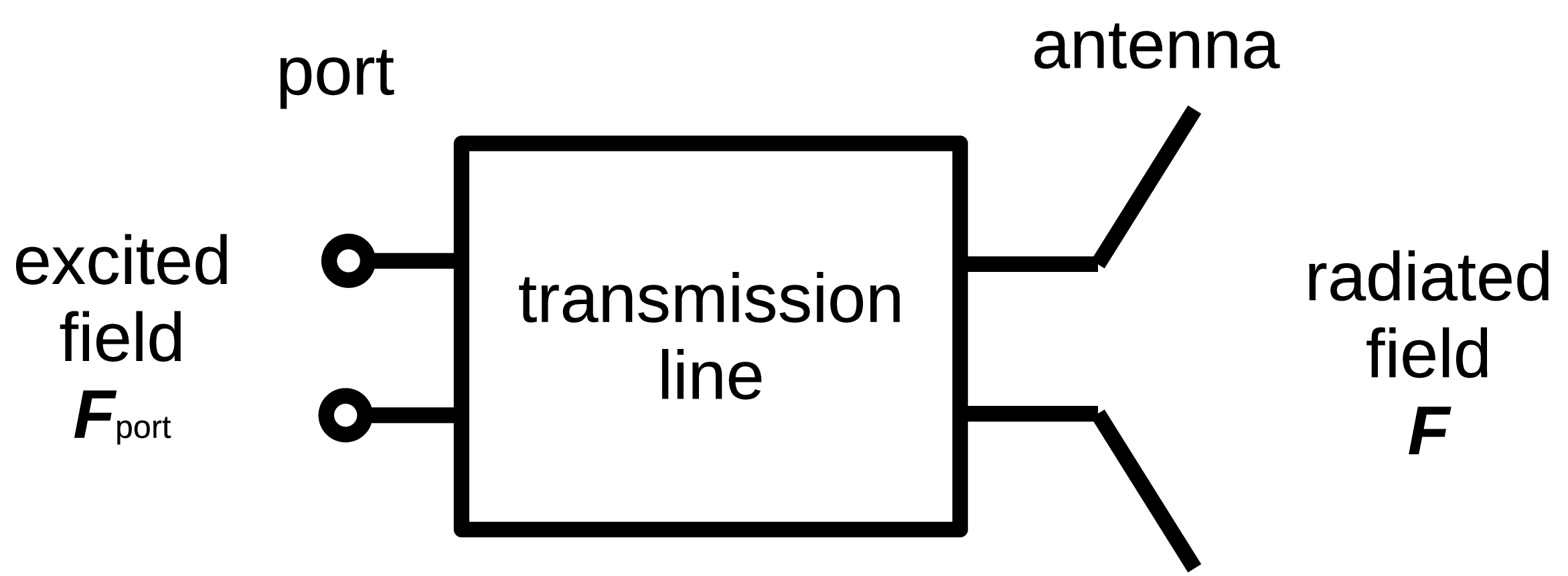} 
 \end{center}
\caption{Conceptual model. An EM field is excited at the port and propagated through a transmission line. An antenna radiates a beam in a linear medium.}\label{fig:concept}
\end{figure}

We suppose that a system of interest consists of a port of a transmission line such as a waveguide, microstrip, and coaxial cable, an antenna to radiate the power transferred from the port, and a linear medium (Fig.~\ref{fig:concept}).
Since a transmission line and an antenna are also a linear component in most cases, we may expect that a radiated field in the medium is given as a superposition of the excited electromagnetic field $ \bm{ F }_\mathrm{ port } \in \mathbb{ C }^p ~ ( p \in \{ 2, 3, 4, 5, 6 \} ) $ at the port,
\begin{align}
    \bm{ \mathcal{ F } } ( \bm{ r }_1 ) = \mathsf{ A }_1 \bm{ F }_\mathrm{ port }, \label{eq:single_point}
\end{align}
where $ \bm{ \mathcal{ F } } \in \mathbb{ C }^q ~ ( q \in \{ 2, 3, 4, 6 \} ) $ is the radiated field in the medium at a point $ \bm{ r }_1 $ and $ \mathsf{ A }_1 
$ the linear operator calculating $ \bm{ \mathcal{ F } } $ from $ \bm{ F }_\mathrm{ port } $. 
There are various ways to describe a field in a transmission line, $ \bm{ F }_\mathrm{ port } $.
TE modes in a waveguide, for instance, have 5 non-zero vector components, and in that case we may assign 5 to the number of $ \bm{ F }_\mathrm{ port } $ components, i.e., $ \bm{ F }_\mathrm{ port } \in \mathbb{ C }^5 $.
We often take an electric field only for a radiated field but can include a magnetic field.
When either an electric field or a magnetic field is taken, the number of EM field components is reduced to 2 for far field and 3 for near field, respectively.
An example of the linear operators is a diffraction integral, which directly and linearly relates a field in a transmission line to a radiated field.
Some cases reduce the linear operator to a simple coefficient matrix, which will appear in the later sections.

When we have $ n $ points of the radiated field, $ \bm{ r }_i ~ ( 1 \leq i \leq n )$, (\ref{eq:single_point}) is developed to
\begin{align}
    \bm{ F } = \mathsf{ A } \bm{ F }_\mathrm{ port }, ~
    \bm{ F } := \begin{bmatrix}
        \bm{ \mathcal{ F } } ( \bm{ r }_1 ) \\
        \bm{ \mathcal{ F } } ( \bm{ r }_2 ) \\
        \vdots \\
        \bm{ \mathcal{ F } } ( \bm{ r }_n )
    \end{bmatrix}, ~
    \mathsf{ A } :=
    \begin{bmatrix}
        \mathsf{ A }_1 \\ 
        \mathsf{ A }_2 \\ 
        \vdots \\ 
        \mathsf{ A }_n
    \end{bmatrix}. \label{eq:model}
\end{align}
When the inverse operator of $ \mathsf{ A } $ is defined, e.g., inverse Fourier transform, we will retrieve $ \bm{F}_\mathrm{port} $.
In the case that $ \mathsf{ A } $ is a matrix, if
the number of points, $ n $, is larger than or equal to $ p/q $, a generalized inverse matrix or pseudo-inverse matrix $ \tilde{ \mathsf{ A } } $, such that $ \mathsf{ A } \tilde{ \mathsf{ A } } \mathsf{ A } = \mathsf{ A } $, can be computed in many cases.

Regarding practical aspects of a beam measurement, a random noise vector $ \bm{ \varepsilon } \in \mathbb{ C }^{ nq } $ should be added to (\ref{eq:model}), i.e.,
\begin{align}
        \bm{ F } = \mathsf{ A } \bm{ F }_\mathrm{ port } + \bm{ \varepsilon }. \label{eq:model_w_noise}
\end{align}
We may take an appropriate noise model for a system of interest.
According to the Gauss-Markov theorem, the excited amplitudes can be estimated with
\begin{align}
    \hat{ \bm{ F } }_\mathrm{ port } := \left( \mathsf{ A }^\dagger \mathsf{ A } \right) ^{ -1 } \mathsf{ A }^\dagger
    \begin{bmatrix}
        \bm{ \mathcal{ F } } ( \bm{ r }_1 ) \\
        \bm{ \mathcal{ F } } ( \bm{ r }_2 ) \\
        \vdots \\
        \bm{ \mathcal{ F } } ( \bm{ r }_n )
    \end{bmatrix},
    \label{eq:f_est}
\end{align}
where $ \dagger $ denotes a transposed and complex-conjugate matrix of a matrix.
If there are no correlations among $ \varepsilon_i $, and the variance $ \sigma $ is common to all $ \varepsilon_i $, then, the variance of $ \hat{ \bm{ F } }_\mathrm{ port } $ is given by
\begin{align}
\label{VM}
    V \left( \hat{ F }_\mathrm{ port } \right) = \left[ \left( \mathsf{ A }^\dagger \mathsf{ A } \right)^{ -1 } \right]_{ ii } \sigma^2,
\end{align}
where the subscript $ ii $ denotes the $ ii $ component of the matrix.
If $ \sigma $ is unknown, it can be estimated from the residuals,
\begin{align}
\label{Sigma2}
	\hat{ \sigma }^2 = \frac{ \lVert \mathsf{ A } \hat{ \bm{ F } }_\mathrm{ port } - \bm{ F } \rVert^2 }{ \nu },
\end{align}
where $ \nu $ is the degrees of freedom given by $ n - \mathrm{ rank } ( \mathsf{ A } ) $.

\subsection{Formalism}
In this paper, we focus on an electric field, adopt a circular waveguide as a transmission line, and evaluate a far field radiated by the system.
We may consider other quantities and systems because the assumptions and the model equation~(\ref{eq:model_w_noise}) in Sect.~\ref{sec:assumptions} hold valid in many cases, which will be confirmed in Section~\ref{sec:demonstration}.

The electric field at a port is easily expressed with the modes defined in a transmission line,
\begin{equation}
\label{CE}
\bm{F}_\mathrm{port} ( x, y ) = \sum_{n} c_{n}\bm{E}_{n} ( x, y ),
\end{equation}
where $ ( x, y ) $ is the coordinates defined in the waveguide port, $\bm{E}_{n} ( x, y ) \in \mathbb{ C }^3 $ is the (normalized) electric field of mode $n$ at the port, and $ c_{n} \in \mathbb{ C } $ is the amplitude of the excited mode.
Once $ c_n $ is specified, the electric field at the port is fully characterized.
The EM wave is propagated through the waveguide changing $ c_{n} $.
Since $ \mathsf{ A }_i \bm{ E }_n ( x, y ) $ is a complex number, referred to as $ z_{ i, n } $, we obtain $ \mathcal{ F } ( \bm{ r }_i ) = \sum_n c_n z_{ i, n } $ and a matrix $ \mathsf{ A } $ instead of an expression with an operator.

Similarly to the field at a port, we may expand a radiated field into spherical harmonics for a broad beam,
\begin{equation}
\label{AE}
\bm{F} ( \theta, \phi ) = \sum_{\ell, m} \left( a_{\ell m }^{(0)} \bar{ \bm{ \mathrm{ e } } }_\mathrm{ pol 0 } + a_{\ell m }^{(1)} \bar{ \bm{ \mathrm{ e } } }_\mathrm{ pol 1 } \right) Y_\ell^m ( \theta, \phi ),
\end{equation}
where $ ( \theta, \phi ) $ is the spherical coordinates on the celestial sphere,
$ \bar{ \bm{ \mathrm{ e } } }_\mathrm{ pol 0 }$ and $ \bar{ \bm{ \mathrm{ e } } }_\mathrm{ pol 1 } $ are the unit vectors that define the orientations of two linear-polarization components, pol0 and pol1, as a function of $ ( \theta, \phi ) $,
$ a_{ \ell m }^{ ( 0 ) } $ and $ a_{ \ell m }^{ ( 1 ) } $ are the expansion coefficients for each polarization,
and $ Y_\ell^m ( \theta, \phi ) $ is the spherical harmonic function.
It is worth mentioning that we do not always expand a radiated field into a series of the spherical harmonics, in particular, for a narrow beam case.

Even when $ \bm{ F }_\mathrm{ port } $ and $ \bm{ F } $ are expanded into a series of various modes, they are still expressed as a linear combination.
Therefore, we can keep the same model equation as in (\ref{eq:model_w_noise}), and introduce a variation of (\ref{eq:model_w_noise}), for example,
\begin{align}
    \begin{split}
        \bm{ a } &= \mathsf{ A } \, \bm{ c } + \bm{ \varepsilon }, \quad \bm{ a } = \mathsf{ A } \, \bm{ F }_\mathrm{ port } + \bm{ \varepsilon }, \quad \mathrm{ and } \quad \bm{ F } = \mathsf{ A } \, \bm{ c } + \bm{ \varepsilon },
    \end{split}    
    \label{eq:model_w_sph}
\end{align}
where $ \bm{ a }
:= \left( a_{ 0 0 }^{(0)}, a_{ 1 -1 }^{(0)}, a_{ 1 0 }^{(0)}, a_{ 1 1 }^{(0)}, \cdots, a_{ \ell_\mathrm{ max } \ell_\mathrm{ max } }^{(0)}, a_{ 0 0 }^{(1)}, \cdots, a_{ \ell_\mathrm{ max } \ell_\mathrm{ max } }^{(1)} \right)^\mathrm{ T }
$ is the expansion coefficient vector with $ 2 ( \ell_\mathrm{ max } + 1 )^2 $ elements, $ \bm{ c } = \left( c_0, \cdots, c_{ n_\mathrm{ max } } \right)^\mathrm{ T } $ is the mode amplitude vector with $ n_\mathrm{ max } $ elements, and, consequently, $ \mathsf{ A } $ associated with $ \bm{ c } $  has the size according to $ \bm{ F }, ~ \bm{ a } $, or $ \bm{ c } $.
We note that there are several kinds of basic function to expand the fields into a small number of coefficients in \eqref{CE} and (\ref{AE}).

A pseudo-inverse matrix can be obtained with various methods such as the QR decomposition, the singular value decomposition (SVD), and the like \cite{ben2003generalized}.
If we perform SVD of $\mathsf{ A }$, that is,
\begin{align}
\mathsf{ A } = \mathsf{ U } \begin{bmatrix}
\mathsf{ \Sigma } & \mathsf{ 0 } \\
\mathsf{ 0 } & \mathsf{ 0 } \\
\end{bmatrix}
\mathsf{ V }^{\dagger},
\label{eq:svd}
\end{align}
the pseudo-inverse matrix $ \tilde{ \mathsf{ A } } $ is given by
\begin{align}
    \tilde{ \mathsf{ A } } = \mathsf{ V }
    \begin{bmatrix}
    	\mathsf{ \Sigma }^{-1} & \mathsf{ 0 } \\
		\mathsf{ 0 } & \mathsf{ 0 } \\
\end{bmatrix}
\mathsf{ U }^{\dagger},
\label{Adag}
\end{align}
where $ \mathsf{ U } $ and $ \mathsf{ V } $ are complex unitary matrices and $ \mathsf{ \Sigma } $ is a $ \rank \mathsf{ A } \times \rank \mathsf{ A } $ real diagonal matrix consisting of the singular values of $ \mathsf{ A } $. When (\ref{eq:svd}) holds, simple calculation connects the matrix in (\ref{eq:f_est}) to (\ref{Adag}),
\begin{align}
    \left( \mathsf{ A }^\dagger \mathsf{ A } \right)^{ -1 } \mathsf{ A }^\dagger = \mathsf{ V }
    \begin{bmatrix}
        \mathsf{ \Sigma }^{ -1 } & \mathsf{ 0 } \\
        \mathsf{ 0 } & \mathsf{ 0 }
    \end{bmatrix}
    \mathsf{ U }^\dagger = \tilde{ \mathsf{ A } }.
\end{align}

\subsection{Construction of matrix $ \mathsf{ A } $}
A matrix $ \mathsf{ A } $ is explicitly needed to deduce the excited field at the port from the radiated field.
It can be obtained with a vector holding a single non-zero value such as $ \bm{ c } = ( 0, \cdots, c_i, \cdots, 0 )^\mathrm{ T } $.
It is extremely hard to make such a vector experimentally, though numerical simulation can easily control the excited field at the port.
Therefore, we may rely on numerical simulation to determine $ \mathsf{ A } $.
For example, we simulate beam patterns $ \bm{ F }_i $ by exciting the $ i $-th mode at the port whose amplitude is unity.
Then, placing the resultant $ \bm{ F }_i $ in columns, we obtain a matrix,
\begin{align}
    \begin{bmatrix}
        \bm{ F }_1 & \bm{ F }_2 & \cdots & \bm{ F }_n
    \end{bmatrix},
\end{align}
which exactly works as $ \mathsf{ A } $ in (\ref{eq:model_w_sph}).

\section{Numerical demonstration} \label{sec:demonstration}
This section addresses how matrix $ \mathsf{ A } $ is prepared and how accurately the method works.
Two cases are demonstrated: an axially-corrugated (AC) horn and an offset Cassegrain telescope with a conical horn.
The radiated field of the AC horn is expanded into a series of the spherical harmonics and evaluated through (\ref{eq:model_w_sph}) with $ \bm{ a } $ and $ \bm{ c } $.
The offset Cassegrain telescope case shows the analysis with a model equation consisting of vectors $ \bm{ F } $ and $ \bm{ c } $ in (\ref{eq:model_w_sph}).

\subsection{Spherical harmonics expansion case: axially-corrugated horn} \label{sec:ac}
\begin{figure}[b]
 \begin{center}
  \includegraphics[width=8cm]{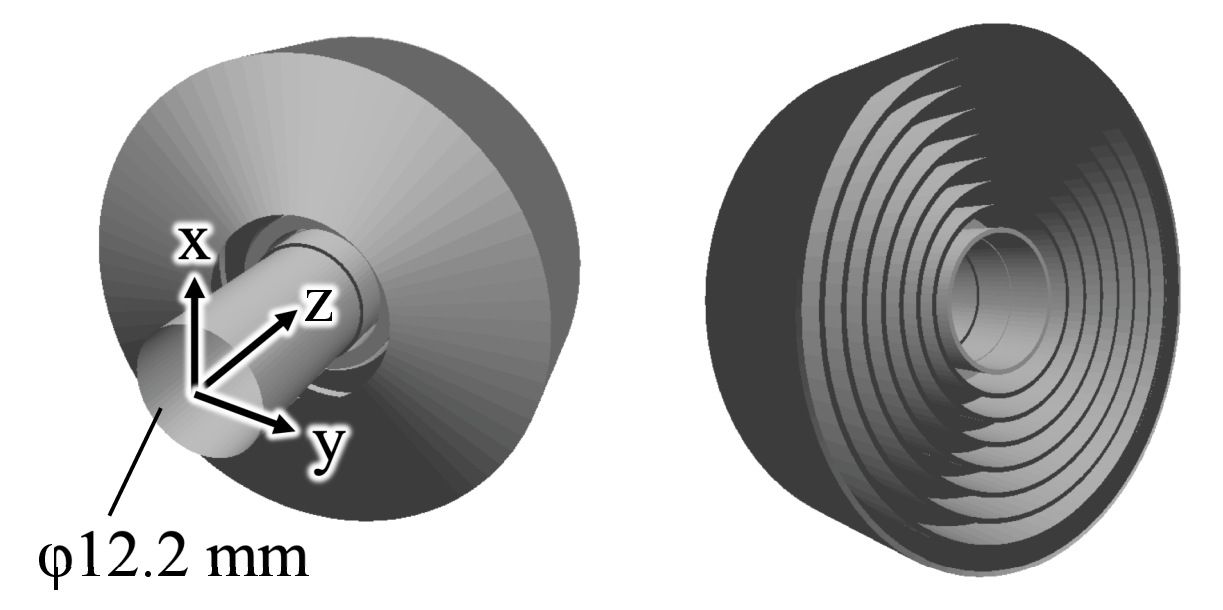} 
 \end{center}
\caption{An axially-corrugated horn with an input waveguide diameter of 12.2\,mm.}\label{fig:model}
\end{figure}

\begin{table}[b]
\centering
\caption{Cutoff frequencies of the circular waveguide with a diameter of 12.2\,mm.}
\label{tab:fc}
\begin{tabular}{cc}
\hline
Mode & $f_\mathrm{cutoff}$/GHz \\
\hline
TE$_{11}^{(0)}$, TE$_{11}^{(1)}$ & 14    \\
TM$_{01}$ & 19   \\
TE$_{21}^{(0)}$, TE$_{21}^{(1)}$ & 24    \\
\hline
\end{tabular}
\end{table}

We prepared an AC horn model (Fig. \ref{fig:model}).
The diameter of the input circular waveguide was 12.2\,mm. The cutoff frequencies of the propagating modes are summarized in Table \ref{tab:fc} \cite{balanis2012advanced}.
The horn was analyzed at 25\,GHz, where $\rm{TE_{11}}$, $\rm{TM_{01}}$ and $\rm{TE_{21}}$ could be propagated, as shown in Table \ref{tab:fc}.
The cutoff frequencies of other modes are higher than 25\,GHz, which allows us to focus on the modes in Table~\ref{tab:fc}.
Since $\rm{TE_{11}}$ and $\rm{TE_{21}}$ modes have 2 orthogonal modes, we distinguish them with the superscript.
Using the MoM simulation implemented in TICRAtools\footnote{https://www.ticra.com/ticratools/}, we simulated the beam patterns $\bm{F}(\theta, \phi)$ on the points defined by the Hierarchical Equal Area isoLatitude Pixelization (HEALPix) \cite{Gorski2005}.
HEALPix pixelates a sphere such that all the pixels have the same solid angle and are arranged with the equal intervals in latitude.
A parameter of $ N_\mathrm{ side } $ determines the number of the pixels.
In the paper, $ N_\mathrm{ side } $ was set 16 and, consequently, the number of the pixels was 3072, the pixel solid angle $ \Omega_\mathrm{ pix } $ was $ 4 \pi / 3072 \approx 4 \times 10^{-3} $\,sr, and the angular resolution $ \theta_\mathrm{ pix } $ was $ \sqrt{ \Omega_\mathrm{ pix } } \approx 3.66^\circ $, respectively.
Each coefficient $ a_{ \ell m } $ represents an angular scale of $ \approx 180^\circ/\ell $, and therefore, we tentatively adopted 15 for the maximum $ \ell $, referred to as $\ell_\mathrm{max}$ hereafter.
The beam patterns were expanded into a series of spherical harmonics functions as shown in \eqref{AE}.
We used a python package, healpy \cite{Zonca2019}, and obtained the spherical harmonic coefficients $a_{\ell m}$ and vector $ \bm{ a }$.

\begin{figure}[t]
\centering
\includegraphics[width=8.5cm]{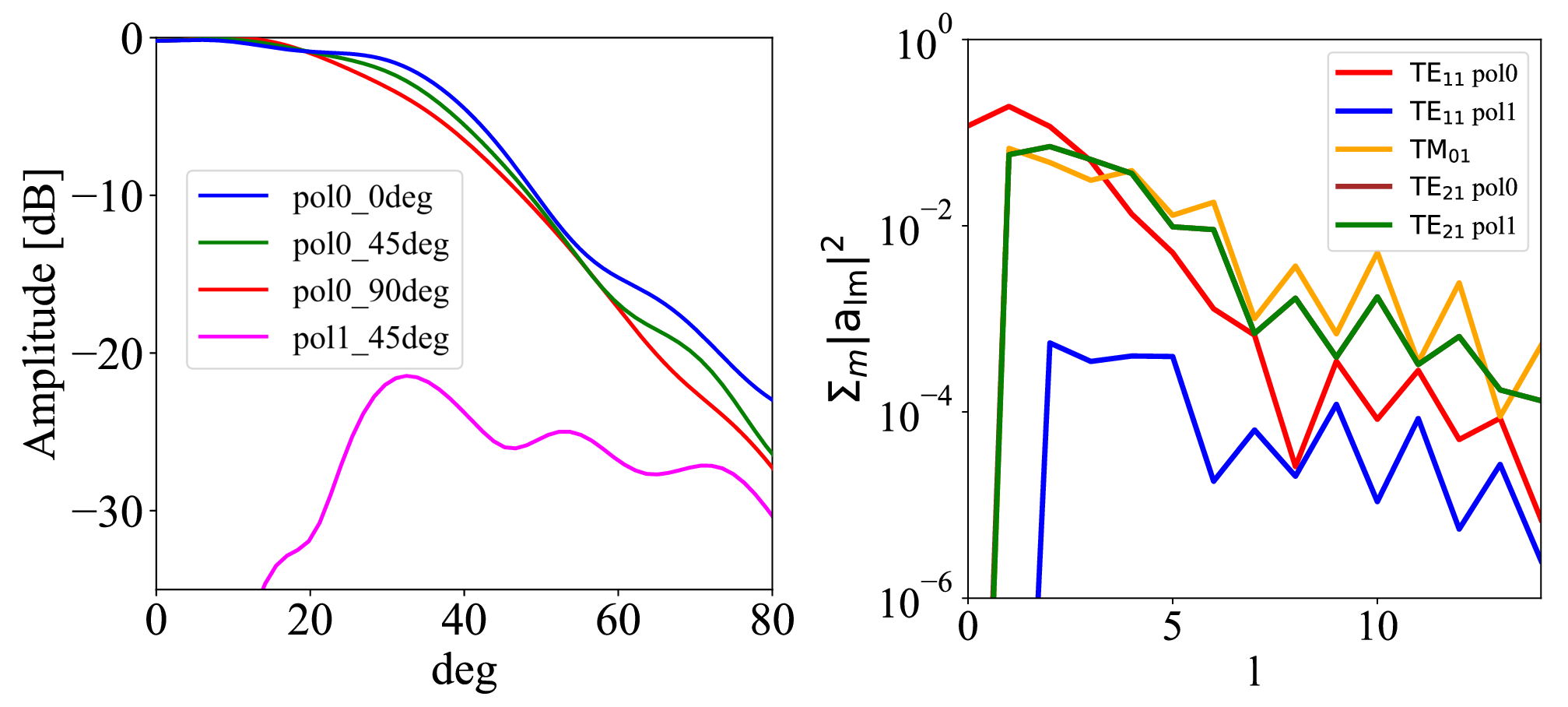}
\caption{Left: Simulated far-field pattern of the axially-corrugated horn with $\rm{TE_{11}^{(0)}}$ mode excited only. Right: The power spectra of the axially-corrugated horn far-field patterns. The coefficients are summed up over $ m $. The TE$_{21}^{(0)}$ and TE$_{21}^{(1)}$ have the identical magnitude, and therefore, two curves are overlapped.}
\label{fig:healpix2}
\centering
  \includegraphics[width=8cm]{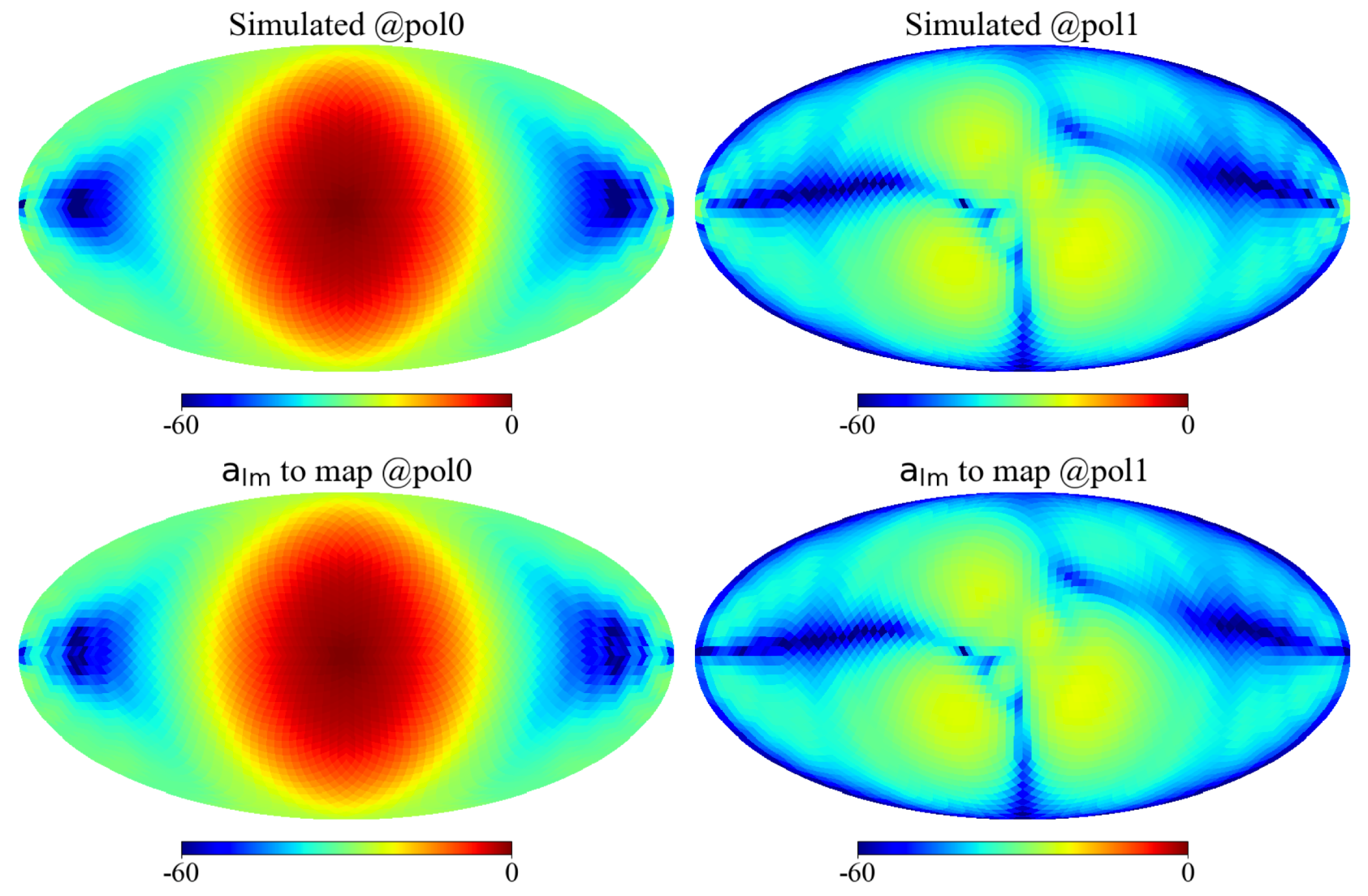} 
\caption{The upper panel shows the original simulated far-field patterns, and the lower the reconstructed beam pattern from the spherical harmonic coefficients. The left column shows pol0, and the right one pol1. All panels are depicted in the Mollweide projection.}
\label{fig:healpix}
\centering
  \includegraphics[width=8cm]{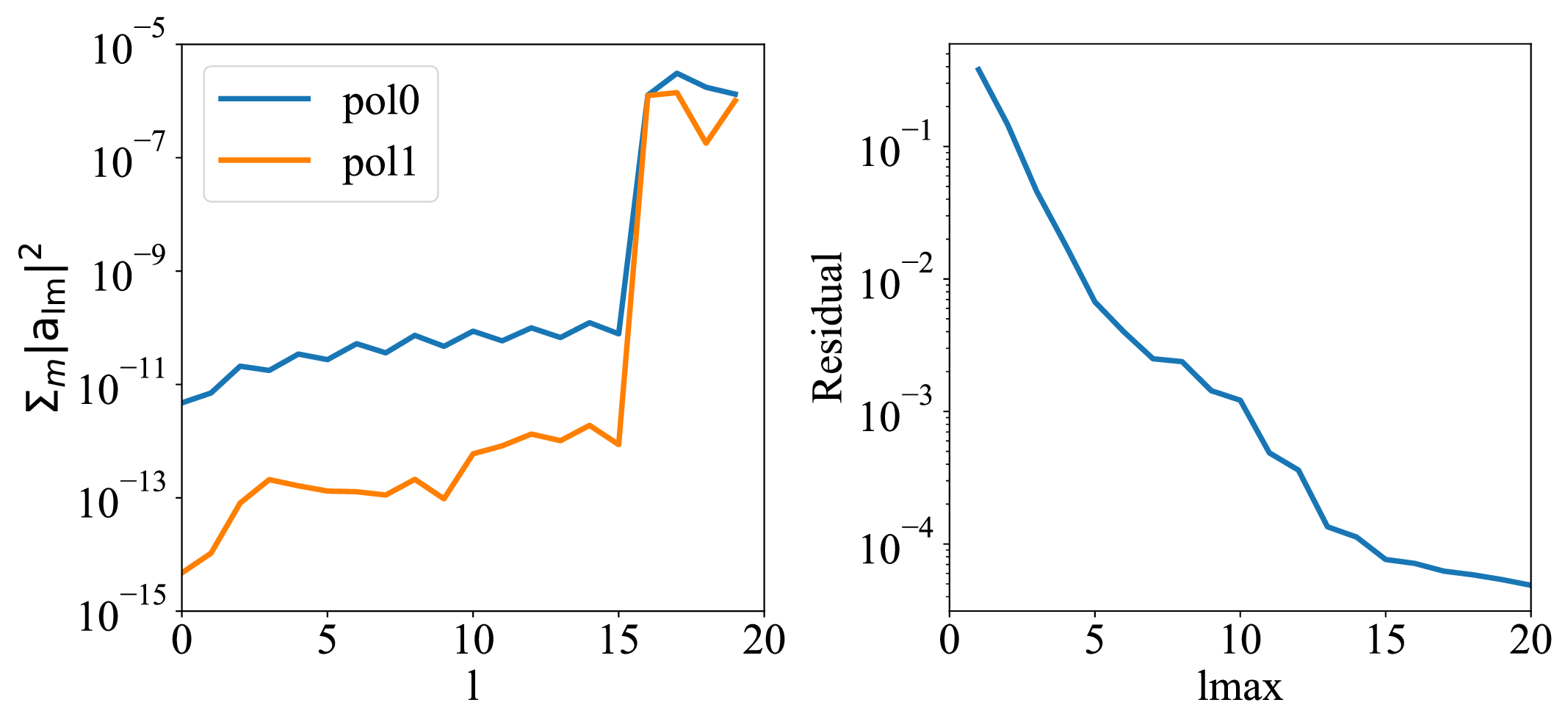} 
\caption{The spherical harmonics expansion coefficients of the residual map between the original and reconstructed beams in Fig.~\ref{fig:healpix}. The coefficients $ a_{ \ell m } $ were integrated over $m$ (left). The residuals depend on $ \ell_\mathrm{ max } $ (right).}
\label{fig:alm_residual}
\end{figure}

The left panel in Fig.~\ref{fig:healpix2} shows the beam cross section and the power spectra when only $ \rm{TE_{11}^{(0)}} $ mode was excited.
Since the matrix $ \mathsf{ A } $ is composed of $ a_{ \ell m } $ for the beams excited by each single mode, the right panel in Fig.~\ref{fig:healpix2} ensures that $ \mathsf{ A } $ does neither have any null columns nor null rows.
It also implies that the power spectra still have the magnitudes of $ 10^{ -4 } $ or so at $ \ell \approx 15 $.

Figure~\ref{fig:healpix} shows the beam pattern with the composite of the five modes listed in Table~\ref{tab:fc} and the reconstructed beam pattern from the coefficients.
The composite beam was obtained by exciting all the five modes at the horn throat in the EM simulation.
We also subtracted the original beam pattern from the reconstructed one and expanded the difference into the spherical harmonics, whose power spectra can be found in the left panel, Fig.~\ref{fig:alm_residual}.
The power spectra at $ \ell \leq 15 $ show the order of $ 10^{ -11 } $, which is a reasonable result because the beam simulation data kept only ten digits when they were outputted from the software.
Therefore, the calculation of $ a_{ \ell m } $ were properly carried out.
In addition, to see how accurately the coefficients reproduce the beam pattern, we define the following figure of merit:
\begin{equation}
    \delta :=\cfrac{\sum_{ \bm{ p } } \left\lVert \sum_{\ell, m} \left( a_{\ell m }^{(0)} \bar{ \bm{ \mathrm{ e } } }_\mathrm{ pol 0 } + a_{\ell m }^{(1)} \bar{ \bm{ \mathrm{ e } } }_\mathrm{ pol 1 } \right) Y_\ell^m ( \bm{ p } ) - \bm{ F } (\bm{ p }) \right\rVert^{2} } {\sum_{ \bm{ p } } \left\lVert \bm{ F } (\bm{ p } ) \right\rVert^{2} },
    \label{eq:residuals}
\end{equation}
where $ \bm{ p } $ represents a point on the sky where the radiated field is calculated.
Figure~\ref{fig:alm_residual} right panel tracks $ \delta $ as a function of $ \ell_\mathrm{ max } $.
The residual $ \delta $ is decreasing monotonically but more slowly than $ \ell < 15 $.
In the practical sense, we conclude that $ \ell_\mathrm{ max } = 15 $ is a sufficient number.

\begin{figure}[t]
    \centering
    \includegraphics[width=5cm]{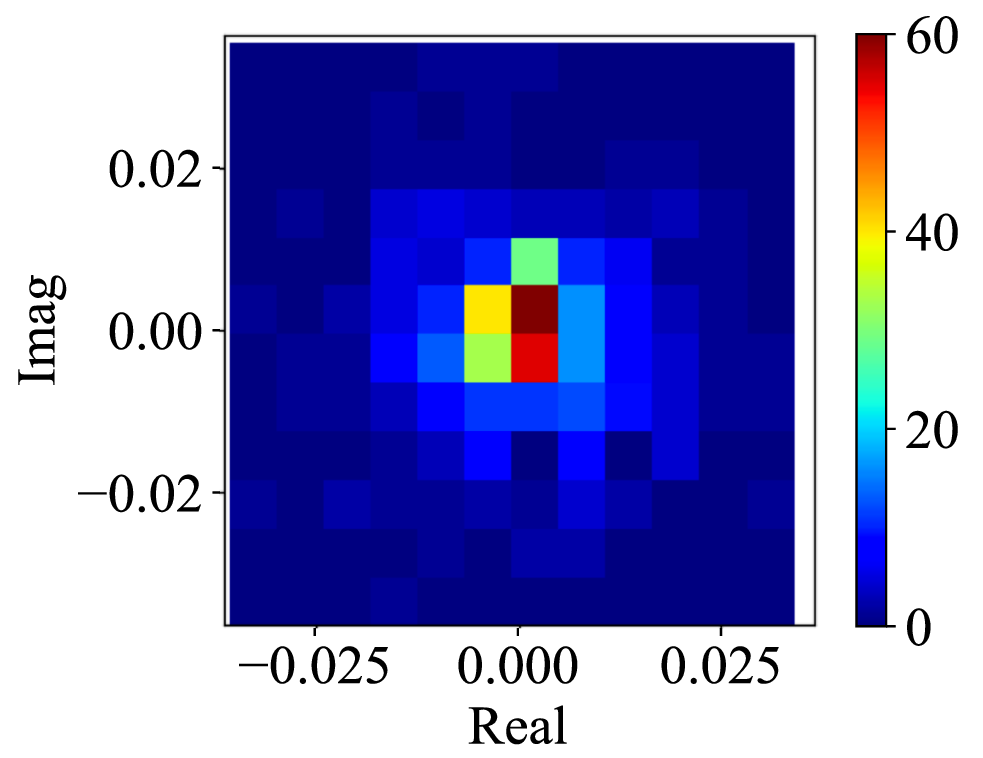} 
    \caption{Histogram of the noise added to the spherical harmonic coefficients $ a_{ \ell m } $. The horizontal and vertical axies denote the real and imaginary part of $ \bm{ \varepsilon } $, respectively. Each pixel is colored according to the number of cases within it.}
    \label{fig:noise}
\end{figure}

\begin{figure}[h]
 \begin{center}
  \includegraphics[width=5cm]{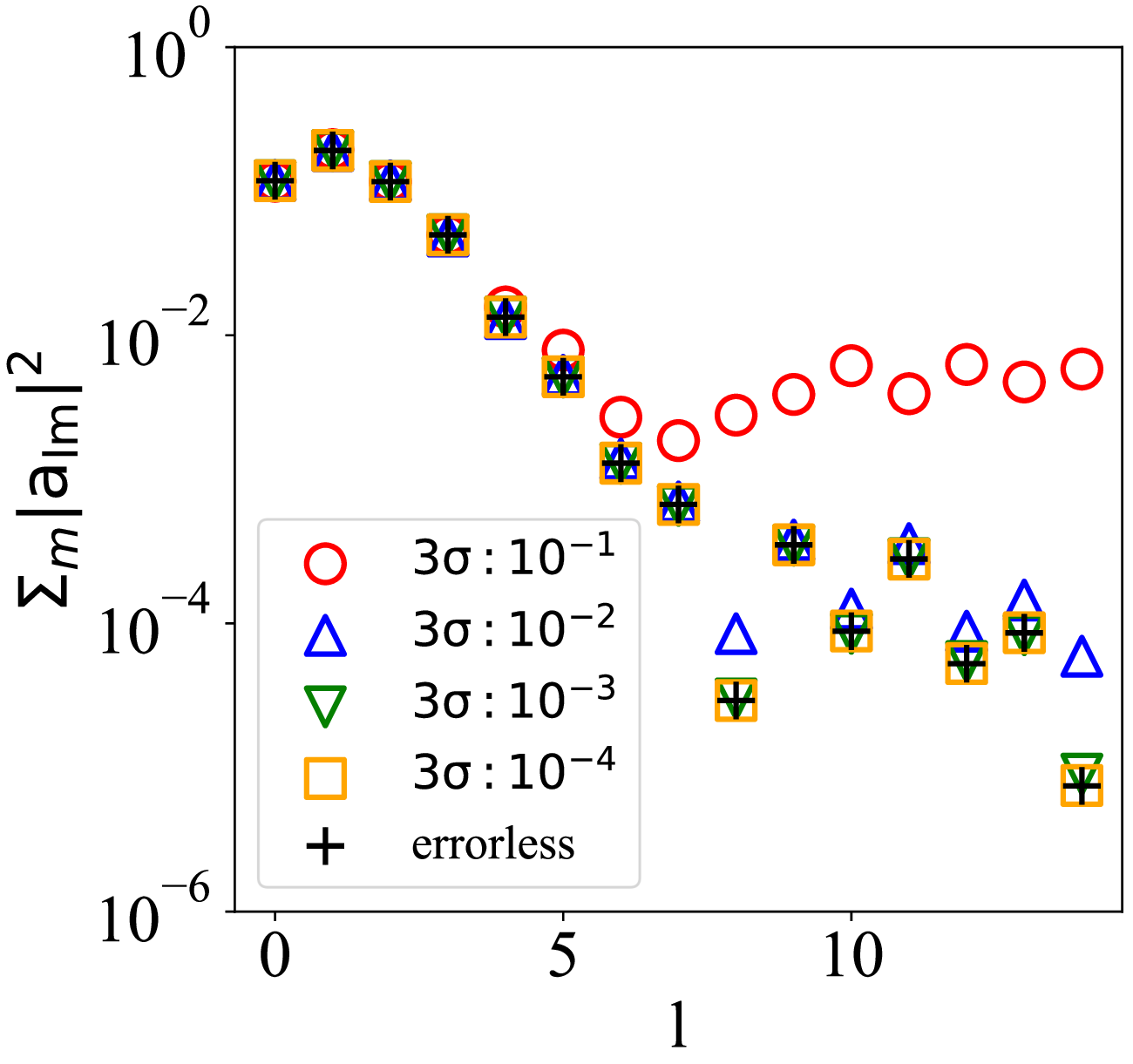} 
 \end{center}
\caption{The power spectra of the composite beam with several error cases. The magnitudes of $3\sigma$ are expressed in the unit of the maximum $|a_{lm}|$.}\label{fig:error}
\end{figure}

After constructing $ \mathsf{ A } $ and $ \tilde{ \mathsf{ A } } $ based on the spherical harmonic coefficient vectors $ \bm{ a } $,
we confirmed that $ \mathsf{ A } $ had full rank, and the product of $ \tilde{ \mathsf{ A } } \mathsf{ A } $ was an identity matrix of size 5.
We also computed the Frobenius norm $ \left\lVert \tilde{ \mathsf{ A } } \mathsf{ A } - \mathbbm{ 1 }_5 \right\rVert_\mathrm{ F } $, resulting in $1.4\times10^{-15}$.
Therefore, we successfully obtained the pseudo-inverse matrix $ \tilde{ \mathsf{ A } } $ with the precision determined by the double-precision floating-point number.

We estimated the mode amplitude vector $\bm{c}$ at the port from the composite beam pattern.
Table~\ref{tab:sh} shows the amplitude and phase of the excited modes, the estimated amplitude, and the difference between them.
It indicates that the amplitude and phase are determined on the order of $10^{-6}$ and $10^{-3}$\,degrees, respectively, for the no error case ($\bm{\varepsilon}=\bm{0}$).
The estimation error of $ 10^{ -6 } $ is probably attributed to the simulation calculation error because
we did not have any significant errors in the spherical harmonics expansion,
$ \left\lVert \tilde{ \mathsf{ A } } \mathsf{ A } - \mathbbm{ 1 }_5 \right\rVert_\mathrm{ F } $ achieved $ O( 10^{-15} ) $,
and we actually confirmed $ \hat{ c } = \tilde{ \mathsf{ A } } \left( \mathsf{ A } c \right) = c + O ( 10^{-15}) $.
Tighter converging criteria for EM simulation could improve the estimation error.

We investigated the effects of errors $ \bm{ \varepsilon } $ on the mode estimation.
Figure~\ref{fig:noise} shows the histogram of the random numbers that we put into the coefficients.
We distributed $ \left| \varepsilon \right|^2 $ according to the chi-squared distribution with one degree of freedom and the argument of $ \varepsilon $ according to a uniform distribution in the range of $ [ 0 : 2 \pi ) $.
Consequently, the real and imaginary parts of $ \bm{ \varepsilon } $ were normally distributed, which were added to the spherical harmonic coefficients of the composite-mode beam.
The mean of the errors was zero, and the standard deviation was prepared with several values.
The three standard deviations of the errors, $ 3 \sigma $, ranged from one-third down to thirty-billionth of the maximum $ | a_{ \ell m } | $.
Figure~\ref{fig:error} shows how the spectra $ \sum_m | a_{ \ell m } |^2 $ are distorted according to the magnitude of the errors.

\begin{figure}[t]
 \begin{center}
  \includegraphics[width=8cm]{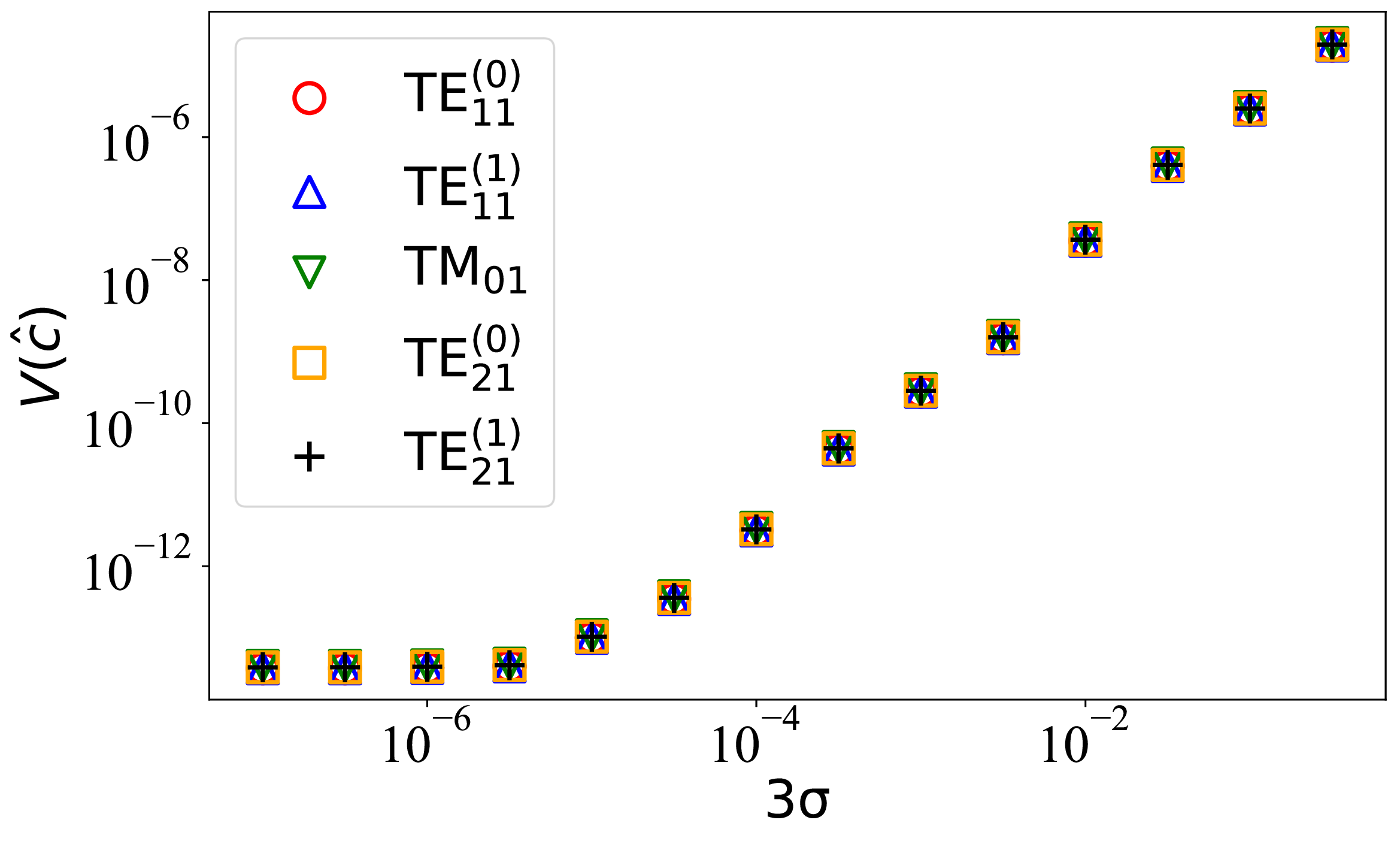} 
 \end{center}
\caption{Variance of the modes $ V( \hat{ c }_i ) $ versus the standard deviation of the error added to $ a _{ \ell m } $ in the unit of the maximum $ | a_{ \ell m } | $.}\label{fig:NASHE}
\end{figure}

Equations~\eqref{VM} and \eqref{Sigma2} were used to calculate the variance of each estimated mode amplitude, $ V( \hat{ c }_i ) $.
Figure~\ref{fig:NASHE} shows that $ V( \hat{ c }_i ) $ decreases as the standard deviation of the errors does.
If the three-sigma deviation of the errors is one-millionth of the maximum $ | a_{ \ell m } | $ or less, systematic or calculation errors dominate the estimation precision.
As Fig.~\ref{fig:error} implies that the deviation $ 3 \sigma / | a_{ \ell m } |_\mathrm{max} \sim 10^{-6} $ will not make any changes in the spectra, it is reasonable that the calculation errors determines the precision.
When systematic errors are dominant, we obtained $ \sqrt{ V( \hat{c}_i ) } \approx 10^{ -7 } $, which agrees the difference in Table~\ref{tab:sh}.


\begin{table*}[t]
\begin{center}
\centering
\caption{Results of the demonstration using the axially-corrugated horn. "Excited mode" represents the amplitude and phase of each mode set up in the electromagnetic field simulation. "Estimation" represents the estimated values by this method.  "Difference" represents the difference between "Excited mode" and "Estimation". "a.u." stands for arbitrary unit.}
\label{tab:sh}
\begin{tabular}{lllllll}
\hline
\multicolumn{1}{c}{}     & \multicolumn{2}{c}{Excited mode}                                          & \multicolumn{2}{c}{Estimation} & \multicolumn{2}{c}{Difference}                                       \\
\hline
\multicolumn{1}{c}{Mode} & \multicolumn{1}{c}{Amplitude in a.u.} & \multicolumn{1}{c}{Phase in deg.} & \multicolumn{1}{c}{Amplitude in a.u.} & \multicolumn{1}{c}{Phase in deg.} & \multicolumn{1}{c}{Amplitude in a.u.} & \multicolumn{1}{c}{Phase in deg.}\\
\hline
\multicolumn{1}{c}{$\rm{TE_{11}^{(0)}}$} & \multicolumn{1}{c}{1}  & \multicolumn{1}{c}{0}  & \multicolumn{1}{c}{1.000000044}             & \multicolumn{1}{c}{$ -4.1 \e{-5} $ }& \multicolumn{1}{c}{ $ 4.4 \e{ -8 } $ }& \multicolumn{1}{c}{ $ -4.1 \e{-5} $ }           \\
\multicolumn{1}{c}{$\rm{TE_{11}^{(1)}}$} & \multicolumn{1}{c}{$ \sqrt{ 0.001 } $ }         & \multicolumn{1}{c}{10}          & \multicolumn{1}{c}{0.031622322}      & \multicolumn{1}{c}{9.999043} & \multicolumn{1}{c}{ $ -4.5 \e{-7 } $ }& \multicolumn{1}{c}{ $ -9.6 \e{-4} $ }   \\
\multicolumn{1}{c}{$\rm{TM_{01}}$} & \multicolumn{1}{c}{$ \sqrt{ 0.001 } $}         & \multicolumn{1}{c}{20}          & \multicolumn{1}{c}{0.031624019}      & \multicolumn{1}{c}{19.999536}& \multicolumn{1}{c}{ $ 1.2 \e{ -6 } $ }& \multicolumn{1}{c}{ $ -4.6 \e{-4} $ }      \\
\multicolumn{1}{c}{$\rm{TE_{21}^{(0)}}$}    & \multicolumn{1}{c}{$ \sqrt{ 0.001 } $}         & \multicolumn{1}{c}{30}          & \multicolumn{1}{c}{0.031622822}       & \multicolumn{1}{c}{30.001900}& \multicolumn{1}{c}{ $ 4.5 \e{-8} $ }& \multicolumn{1}{c}{ $ 1.9 \e{ -3 } $}     \\
\multicolumn{1}{c}{$\rm{TE_{21}^{(1)}}$}    & \multicolumn{1}{c}{$ \sqrt{ 0.001 } $}         & \multicolumn{1}{c}{40}          & \multicolumn{1}{c}{0.031623059}      & \multicolumn{1}{c}{40.004590} & \multicolumn{1}{c}{ $ 2.8 \e{ -7 } $ }& \multicolumn{1}{c}{ $ 4.6 \e{-3} $}     \\
\hline
\end{tabular}
\end{center}
\end{table*}

\subsection{Plane wave expansion case: offset Cassegrain telescope}
\label{B}
To expand a broad beam, such as that of an AC horn, into a series of spherical harmonic functions, $ \ell_\mathrm{ max } $ does not necessarily need a large number.
The case in Sect.~\ref{sec:ac} needs 15 only, because the half width of half maximum power of the beam is 45 degrees, and the spherical harmonic function with $ \ell = 15 $ represents an angular scale of $ \theta_\mathrm{ pix } \approx 4 $ degrees.
Let us take an example of the beam propagated through a large-aperture system. 
Such a beam is so narrow that the spherical harmonic function with very large $ \ell $ must be employed to accurately model it.
Therefore, in the case of a narrow beam, it can be better to leave a far-field radiated pattern as it is. 

\begin{figure}[t]
 \begin{center}
 \includegraphics[width=8cm]{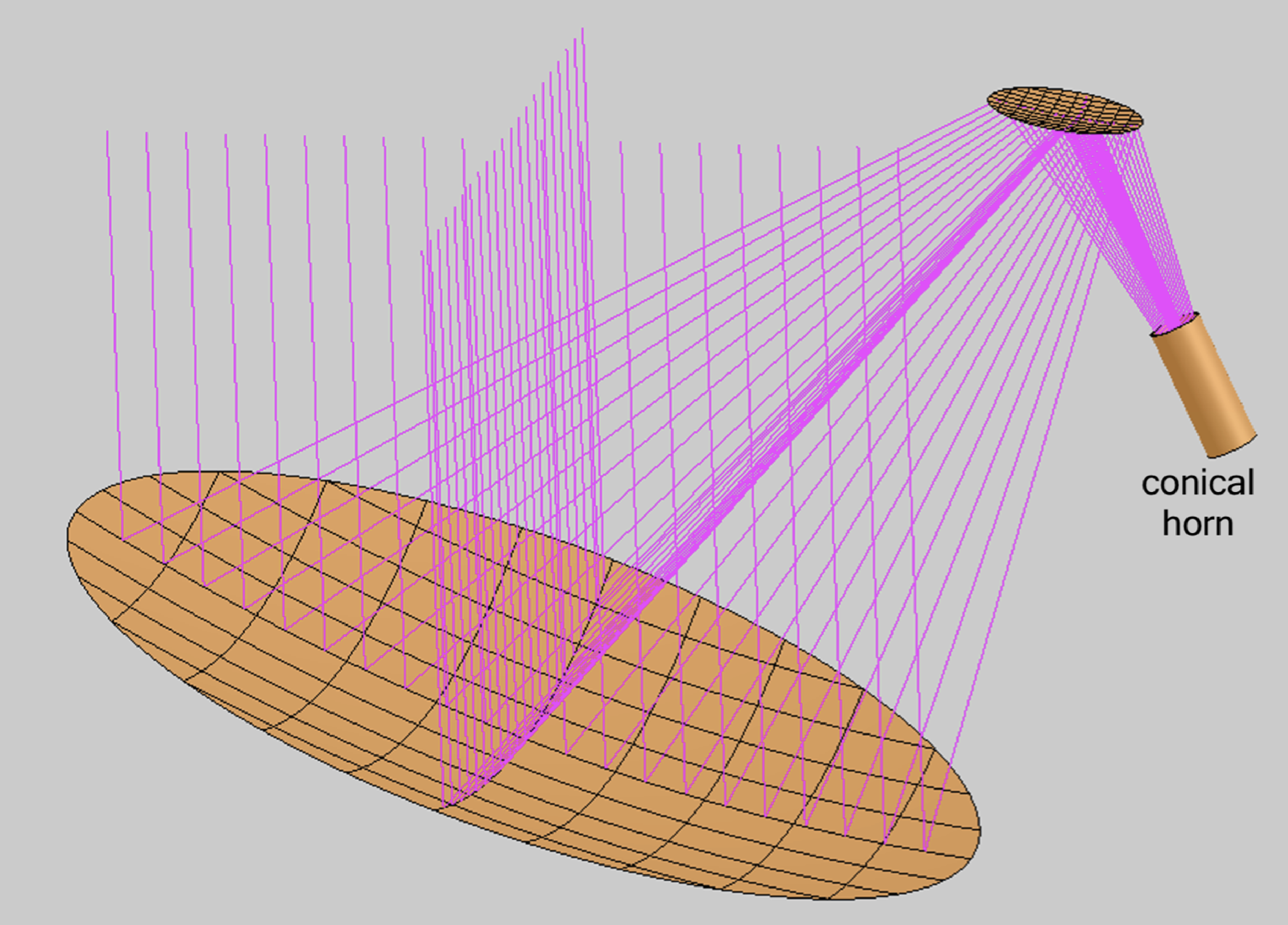} 
\end{center}
\caption{An offset Cassegrain antenna for demonstration.}\label{fig:PWE_model2}
 \begin{center}
  \includegraphics[width=8.5cm]{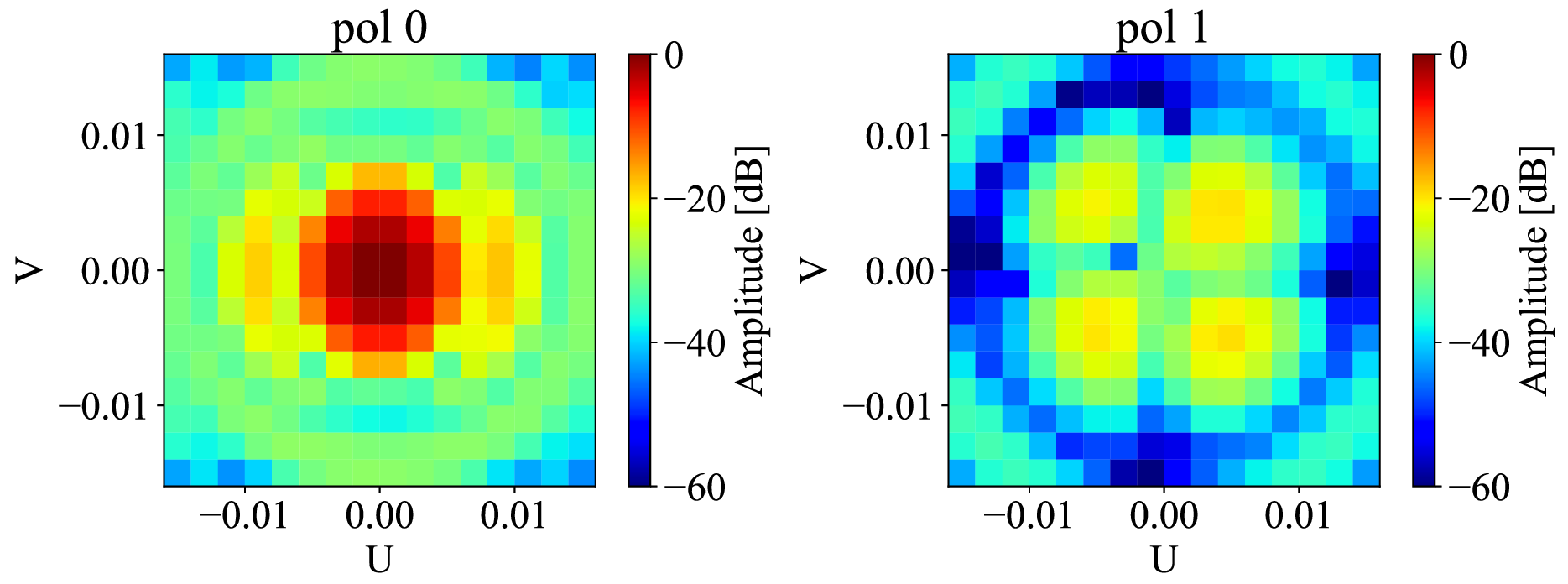} 
 \end{center}
\caption{Radiated beam patterns of the offset Cassegrain antenna as a far field. The left panel shows pol0 and the right pol1. The electric field amplitudes were processed with the method.}\label{fig:PWE_beam}
\end{figure}

An offset Cassegrain antenna with an aperture diameter of 1.2\,m was prepared, as shown in Fig. \ref{fig:PWE_model2}.
The sub-reflector diameter was 200\,mm by 175\,mm.
The simulation frequency was 50\,GHz.
A conical horn was used as the feed, and the 5 waveguide modes in Table~\ref{tab:fc} were excited at 
its throat. 
The conical horn part was simulated with MoM, and then, the secondary and primary reflectors of the offset Cassegrain telescope were computed with Physical Optics simulation.
The beams radiated from the antenna were computed at infinity as a function of $ u := \sin \theta \cos \phi $ and $ v := \sin \theta \sin \phi $, where $ \theta $ and $ \phi $ are the polar and azimuth angles of the spherical coordinates, respectively.
The number of sampling points was $16\times16$ (Fig.~\ref{fig:PWE_beam}).
Both $u$ and $v$ ranged from $-0.015$ to 0.015.
The electric field vectors at each pixel are equivalent to the coefficients of the plane wave expansion.
Finally, we obtained five beams corresponding to each mode and one beam radiated by the composite of the five modes. 
After constructing matrix $ \mathsf{ A } $ from the electric fields of each mode, we confirmed $ \lVert \tilde{ \mathsf{ A } } \mathsf{ A } - \mathbbm{ 1 }_5 \rVert_\mathrm{ F } \approx 1.6\times10^{-15} $ and obtained the pseudo-inverse $ \tilde{ \mathsf{ A } } $ correctly.

\begin{table*}[t]
\begin{center}
\centering
\caption{Results of the demonstration using the offset Cassegrain antenna. "Excited mode" represents the amplitude and phase of each mode set up in the electromagnetic field simulation. "Estimation" represents the estimated values by this method.  "Difference" represents the difference between "Excited mode" and "Estimation". "a.u." stands for arbitrary unit.}
\label{tab:sh2}
\begin{tabular}{lllllll}
\hline
\multicolumn{1}{c}{}     & \multicolumn{2}{c}{Excited mode}                                          & \multicolumn{2}{c}{Estimation} & \multicolumn{2}{c}{Difference}                                       \\
\hline
\multicolumn{1}{c}{Mode} & \multicolumn{1}{c}{Amplitude in a.u.} & \multicolumn{1}{c}{Phase in deg.} & \multicolumn{1}{c}{Amplitude in a.u.} & \multicolumn{1}{c}{Phase in deg.} & \multicolumn{1}{c}{Amplitude in a.u.} & \multicolumn{1}{c}{Phase in deg.}\\
\hline
\multicolumn{1}{c}{$\rm{TE_{11}^{(0)}}$} & \multicolumn{1}{c}{1}  & \multicolumn{1}{c}{0}  & \multicolumn{1}{c}{1.000000091}             & \multicolumn{1}{c}{ $- 1.4 \e{-5} $ }& \multicolumn{1}{c}{ $ 9.1 \e{-8} $ }& \multicolumn{1}{c}{ $ -1.4 \e{-5} $ }           \\
\multicolumn{1}{c}{$\rm{TE_{11}^{(1)}}$} & \multicolumn{1}{c}{$ \sqrt{ 0.001 } $ }         & \multicolumn{1}{c}{10}          & \multicolumn{1}{c}{0.031622843}      & \multicolumn{1}{c}{10.002575} & \multicolumn{1}{c}{ $ 6.7 \e{-8} $ }& \multicolumn{1}{c}{$ 2.6 \e{-3} $ }   \\
\multicolumn{1}{c}{$\rm{TM_{01}}$} & \multicolumn{1}{c}{$ \sqrt{ 0.001 } $}         & \multicolumn{1}{c}{20}          & \multicolumn{1}{c}{0.031623303}      & \multicolumn{1}{c}{19.998905}& \multicolumn{1}{c}{ $ 5.3 \e{-7} $ }& \multicolumn{1}{c}{$-1.1 \e{-3} $}      \\
\multicolumn{1}{c}{$\rm{TE_{21}^{(0)}}$}    & \multicolumn{1}{c}{$ \sqrt{ 0.001 } $}         & \multicolumn{1}{c}{30}          & \multicolumn{1}{c}{0.031623763}       & \multicolumn{1}{c}{30.003552}& \multicolumn{1}{c}{ $ 9.9 \e{-7}  $ }& \multicolumn{1}{c}{ $ 3.6 \e{-3} $}     \\
\multicolumn{1}{c}{$\rm{TE_{21}^{(1)}}$}    & \multicolumn{1}{c}{$ \sqrt{ 0.001 } $}         & \multicolumn{1}{c}{40}          & \multicolumn{1}{c}{0.031624924}      & \multicolumn{1}{c}{40.003813} & \multicolumn{1}{c}{ $ 2.1 \e{-6} $ }& \multicolumn{1}{c}{ $ 3.8 \e{-3}$ }     \\
\hline
\end{tabular}
\end{center}
\end{table*}

Using the obtained pseudo-inverse $ \tilde{ \mathsf{ A } } $, we deduced the coefficient vector of the modes, $ \bm{ c } $, at the conical horn port.
The differences between the estimated values and the excited amplitude in the simulation are shown in Table \ref{tab:sh2}.
The amplitude was determined on the order of $10^{-6}$ with respect to the beam peak and the phase on the order of $10^{-3}$ degrees, respectively.
It is worth noting that we achieve the mode amplitude estimation for the Cassegrain antenna with a similar precision to that for the AC horn case.
In other words, the method will provide us with robust estimation regardless of a specific component or system unless the assumptions in Sect.~\ref{sec:assumptions} are broken.

\begin{figure}[t]
 \begin{center}
  \includegraphics[width=8cm]{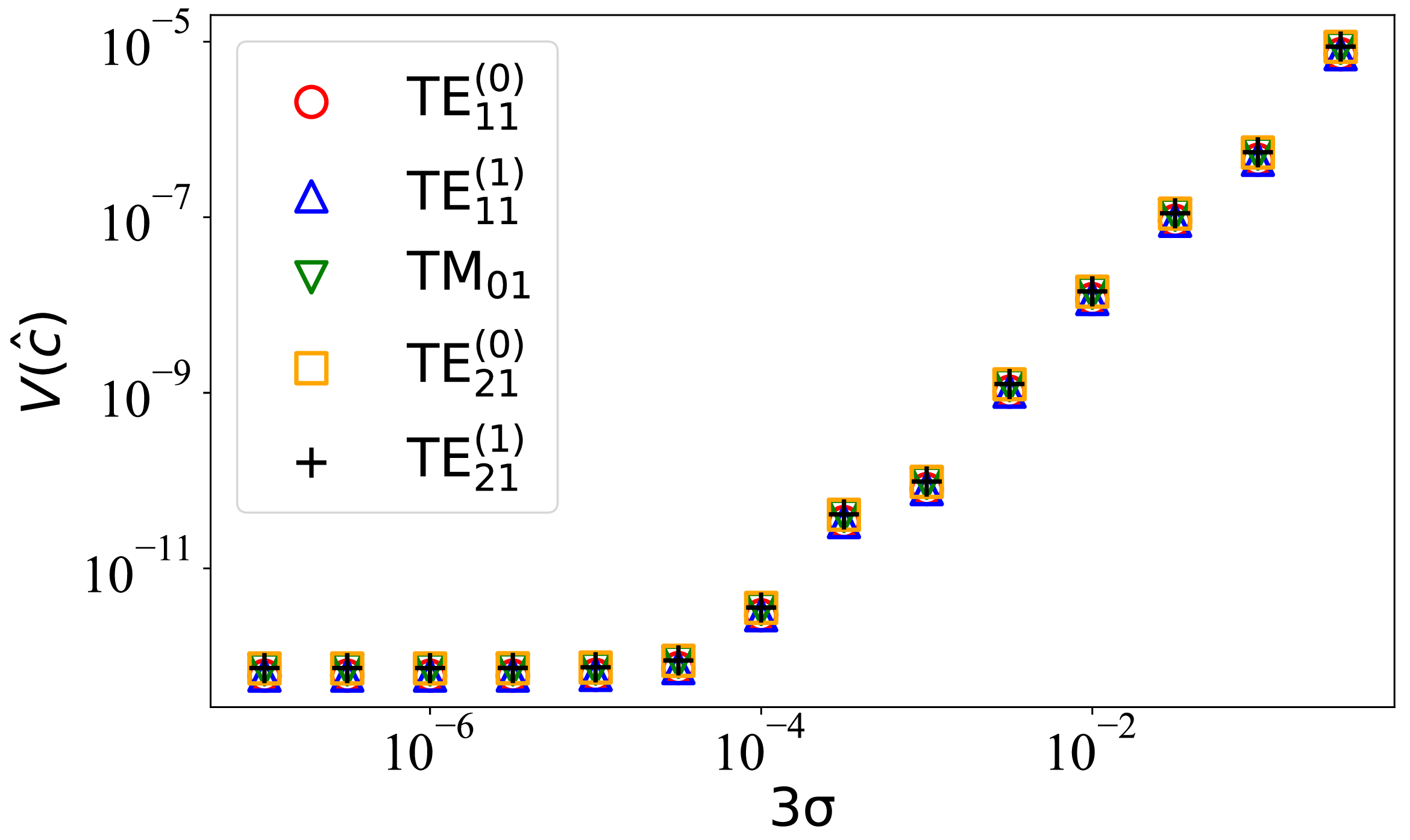} 
 \end{center}
\caption{Variance of the modes $ V(\hat{ c }_i ) $ versus the standard deviation of the errors added to the simulated beam, $ \sigma $.}\label{fig:PWE_model}
\end{figure}

We investigated the effect of the errors on the determination of the mode coefficients in the same manner as in Sect.~\ref{sec:ac}.
Normally distributed complex random errors were added to the simulated beam, and the mode coefficients were estimated as well.
Equations~\eqref{VM} and \eqref{Sigma2} were used to calculate the variance of each mode.
Figure~\ref{fig:PWE_model} shows the smaller variance $ V(\hat{c}_i) $ with the standard deviation of the errors, $\sigma$, decreasing.
When $ 3 \sigma < 10^{ -4 } $ with respect to the beam peak amplitude, calculation errors dominate the variance, the magnitude of which, $ \sqrt{ V( \hat{c}_i ) } \approx 10^{-6} $, is consistent with the differences in Table~\ref{tab:sh2}.

\section{Discussion} \label{sec:discussion}

\subsection{Application to other systems}
The two cases in Sect.~\ref{sec:demonstration} indicate that we may apply the method to any systems.
This is because we only have introduced the simple assumption in Sect.~\ref{sec:assumptions}, that is, linearity, and also, we can rely on the linear algebra.
Those two facts ensure that the method provides the estimation of excited amplitudes with a good precision of $ 10^{ -6 } $ or less with respect to the maximum amplitude, independently of a specific component configuration.
The precision of $ 10^{ -6 }$ would be determined by various causes: computation convergence in EM simulation, the number of digits transferred from simulation, the number of modes used for the far-field expansion, inherent errors in matrix calculation algorithms, and the like.
As long as linearity holds, we may adopt any expressions for electromagnetic fields as shown in (\ref{eq:model_w_sph}), according to our preference.
In fact, judging from
the beam widths, we have demonstrated which quantity to be employed for calculation; the spherical harmonic coefficients or the far field itself (the plane wave expansion coefficients).
This nature would make it easier to apply the method to other systems.

\subsection{Diagnostic of feed alignment}
\begin{figure}[t]
 \begin{center}
  \includegraphics[width=8cm]{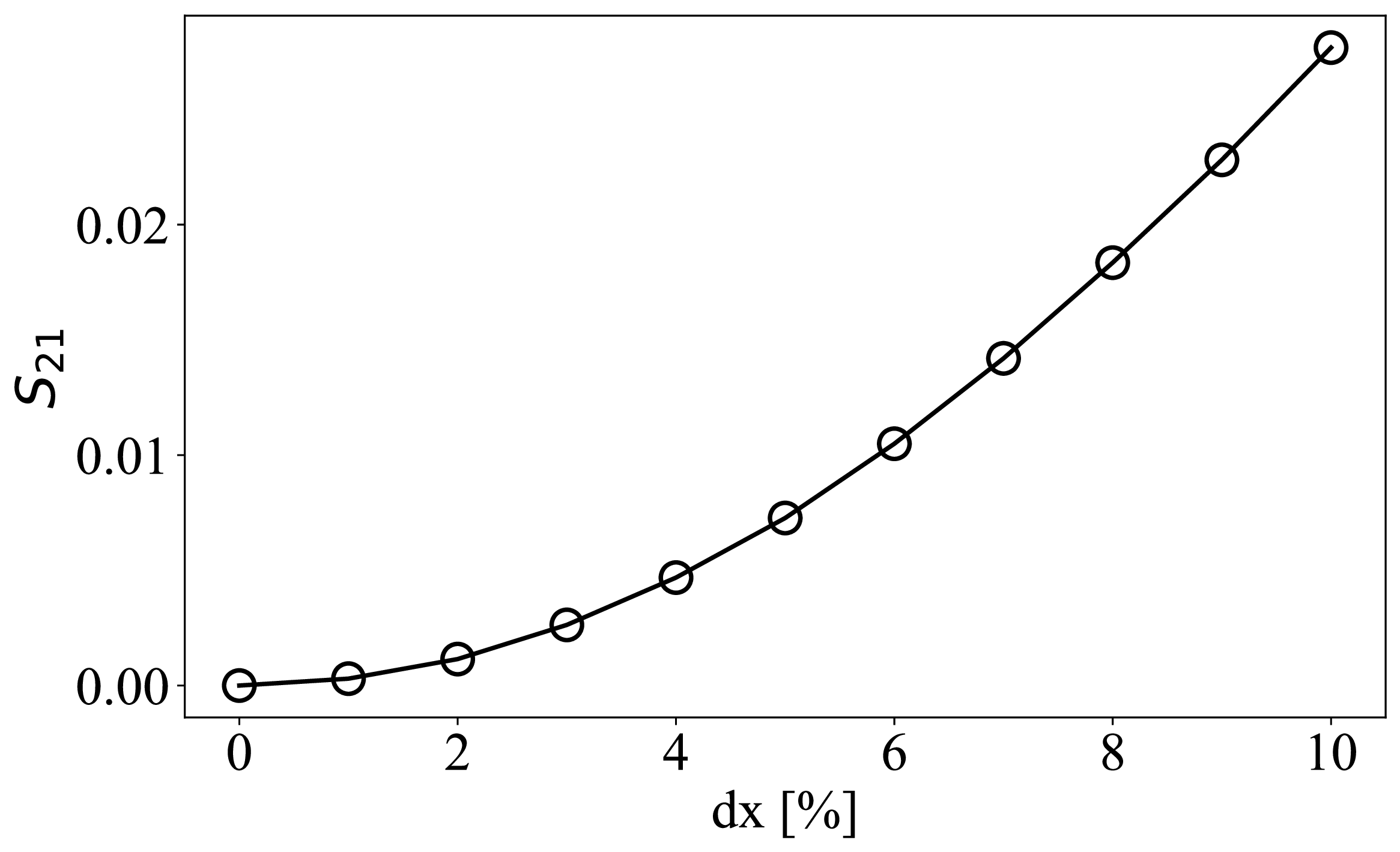} 
 \end{center}
\caption{Simulation of higher-order modes on waveguide discontinuity. The power of $\rm{TE_{21}^{(1)}}$ mode at a waveguide which has offset in the direction of the fundamental-mode electric vector.}
\label{fig:align}
\end{figure}

The proposed method is useful for the alignment diagnostic between two components.
Take an example: a beam measurement system that employs a feed horn (e.g., \cite{7419273}) associated with a circular-to-rectangular waveguide transition.
The horn and the transition should be secured
for measurement accuracy. 
Their displacement 
causes the higher-order modes at the interface plane. 
The waveguide discontinuity distorts
the beam pattern of the 
horn and degrades the measurement accuracy. 
We can obtain the mode coefficients from the radiated field with this method.
In addition to the mode deduction, we can compute how much the higher order modes occur at the interface.
\cite{Ruiz-Cruz10} analyzed a general waveguide step with Mode-Matching Methods, 
which enables us to calculate
the higher-order mode coefficients as a function of displacement.
Fig.~\ref{fig:align} shows the power of TE$_{21}^{(1)}$ as a function of a lateral waveguide displacement.
If we find 1\% of TE$_{21}^{(1)}$ from a beam pattern measurement, the waveguides could have a lateral shift of 5.5\% with respect to the waveguide diameter.
Therefore, 
this estimation method helps us
minimize the mismatch in combination with the predicted magnitudes of higher-order modes.

\subsection{Optimization of excited modes at a feed point}

Let us revisit the process of designing a feed horn. We usually have various requirements regarding a feed. One of the most commonplace ones is the cross-polarization level of a feed beam. We prefer as low cross-polarization level as possible in most applications. To lessen the cross-polarization power, we usually have two ways for our preference. One is to search for a geometrical shape of a feed to achieve lower cross-polarization. The other is to adjust excited modes at a feed point.

Take the corrugated horn in Section~\ref{sec:ac} as an example and excite five modes at its throat.
Then, we have five degrees of freedom; in other words, we may impose up to five conditions on the excited modes.
The cross-polarization level in the radiated field (Fig.~\ref{fig:healpix2}) were dominated by $ a_{ 2, \pm 2 } $.
Setting smaller values of $ a^\prime_{ 2, \pm 2 } $ than $ a_{ 2, \pm 2 } $, we may have two equations,
\begin{align}
a^\prime_{ 2, 2 } & = \sum_{m=1}^5 \mathsf{ A }_{2,2,\mathrm{mode}\,m} c_{\mathrm{mode}\,m}, \\
a^\prime_{ 2, -2 } & = \sum_{m=1}^5 \mathsf{ A }_{2,-2,\mathrm{mode}\,m} c_{\mathrm{mode}\,m},
\end{align}
where $ \mathsf{ A }_{2, \pm 2,\mathrm{mode}\,m} $ are the matrix elements that determine $ a_{ 2, \pm 2 } $, and $ c_{\mathrm{mode}\,m} $ is the amplitude of the excited mode $ m $.
Under these conditions, the degrees of freedom decrease from five to three.
The smaller parameter space makes it easy that we determine the ratios among the mode coefficients $ a_{\ell,m} $.
Thus, the proposed method can be a powerful tool to adjust and determine excited modes at a feed point.

\subsection{Upgrading an existing system}
\begin{figure}[t]
 \begin{center}
  \includegraphics[width=6cm]{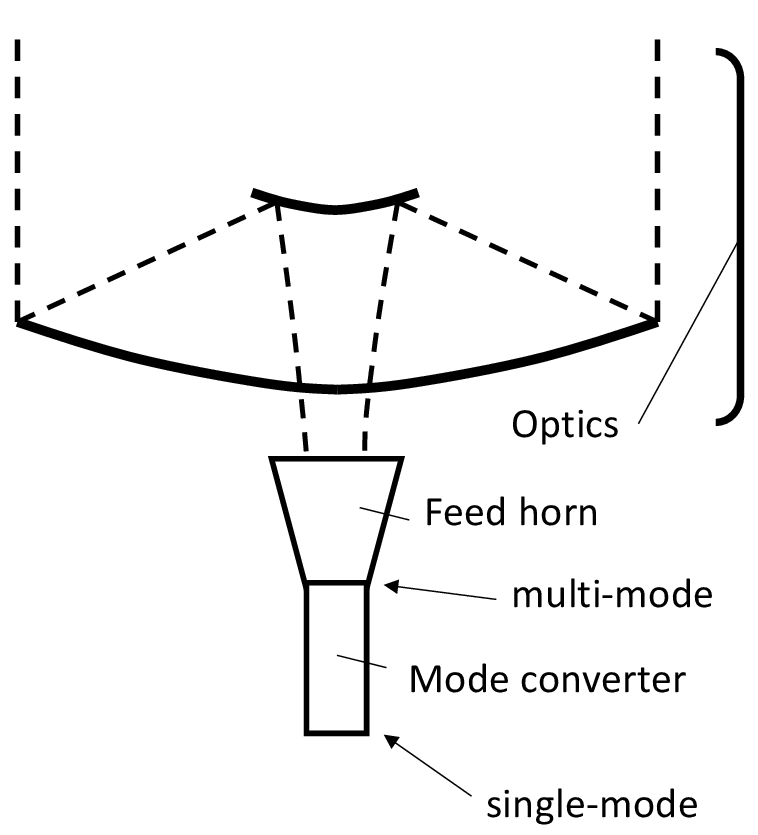} 
 \end{center}
\caption{Concept drawing of upgrading an existing system consisting of a large aperture antenna, relaying optical system, and a feed horn by an additional waveguide mode converter attached to the feed horn throat.}\label{fig:dis}
\end{figure}

Suppose that we attempt to upgrade an existing system consisting of a large aperture antenna, relaying optical system, and a feed horn in Fig.~\ref{fig:dis} by an additional waveguide mode converter attached to the feed horn throat.
The boundary condition of the upgrading is that the existing part are fixed, and the additional mode converter shape should be optimized.
Setting proper requirements on the far field radiated from the existing system and making a matrix connecting the radiated field and the electromagnetic field at the feed horn throat, we can calculate the mode coefficients at the throat to meet the far-field requirements as much as possible, with the proposed method.
Based on the obtained mode coefficients, we may choose the initial parameters of the mode converter which achieve the desired EM field at the mode converter edge by itself when a fundamental mode is excited at the other edge.
Thus, the optimization including both the existing system and the mode converter is started with the initial parameters in order to satisfy the requirements as the whole system.
This strategy would reduce the duration and cost of optimization.


\subsection{Limitation of the method}
As demonstrated in Sects.~\ref{sec:ac} and \ref{B}, random errors on the radiated field degrade the mode-deduction precision.
Figs.~\ref{fig:NASHE} and \ref{fig:PWE_model} imply how accurate measurement of a beam is needed to capture the characteristics of excited modes. For example, \cite{Villiers2022} measured the primary beam of MeerKAT with an S/N of 100 at maximum, which will enables us to deduce the mode amplitudes at the source with a precision of $\sim$$10^{-3}$ (Fig. \ref{fig:PWE_model}) with the method.
We also performed the same analysis in Sects.~\ref{sec:ac} and \ref{B} with a zero-mean uniform distribution instead of a Gaussian one.
We obtain the same result as in Figs.~\ref{fig:NASHE} and \ref{fig:PWE_model} and concluded that the method is less dependent on the noise distribution type.

Regarding non-zero mean noises, there are several causes that we will see.
In terms of making a matrix $ \mathsf{ A } $, we relied on numerical simulation.
If there is any discrepancy of a simulation model from an actual system, the obtained matrix $ \mathsf{ A } $ cannot accurately characterize it.
If we carry out the mode deduction from the beam measurement with an erroneous matrix, the deduced mode amplitudes would be biased and lose accurate values.
Other factors that give rise to systematic errors are, for instance, the probe-horn compensation \cite{Joy1978,Paris1978}, misalignment in a measurement system, and biases such as gain variation.
Those systematic errors would be identified if we find any excess of the predicted higher-order modes, e.g., Fig.~\ref{fig:dis}.
In a practical sense, we would see some systematic errors on the mode deduction.
An experimental demonstration is a future work.


\section{Conclusions}
Assuming linearity in a system consisting of a transmission line and an antenna, we have developed and demonstrated the method to deduce the excited amplitude at the transmission line port from the radiated field by the antenna.
Thanks to the simple assumption that generally holds and the linear algebra, the method can be applied to any linear systems.
As a result, the two cases have clearly shown that the method works well independently of a specific configuration.
The details are summarized below:
   \begin{enumerate}
      \item We prepared the axially-corrugated horn and expanded the far-field simulation data into a series of the spherical harmonics up to $ \ell \leq 15 $.
      Then, we constructed the matrix connecting the spherical harmonic coefficients and the excited mode coefficients at the throat of the horn and computed its pseudo-inverse.
      The excited mode coefficients were  estimated on the order of $10^{-6}$ with respect to the maximum mode coefficients. The mode coefficient phases were also done with a precision of $10^{-3}$ degrees.
      \item For the offset Cassegrain antenna illuminated by a conical horn, we made an attempt at the demonstration similar to the axially-corrugated horn case. We prepared the far-field data in the limited range of $ u $ and $ v $ and constructed the matrix connecting the electric field vectors and the excited modes at the throat of the conical horn. The mode coefficients can be determined in the order of $10^{-6}$ in amplitude and $10^{-3}$ degrees in phase, with respect to the maximum mode amplitude.
      \item Simulations of adding errors to the coefficients or the far-field show that the precision of the estimation depends on the errors as expected. In addition, when the extremely small errors were added, we observed the systematic errors by the calculation. In our cases, the calculation errors dominated the mode-estimation precision up to $ \sqrt{V(\hat{c}_i)} \sim 10^{-6} $ with respect to the maximum mode amplitude, and the random errors did at a higher level of that.
   \end{enumerate}
Since the demonstrated method can be employed for a general case, we may consider various applications, for example, diagnostic of feed alignment and feed design.

\section*{Acknowledgment}
The authors are grateful to Shin'ichiro Asayama for useful suggestions. We would like to thank Mattieu S. de Villiers, Mariet Venter, and Adriaan Peens-Hough for useful discussions. This work was supported by JST, the establishment of university fellowships towards the creation of science technology innovation, Grant Number JPMJFS 2138.

\ifCLASSOPTIONcaptionsoff
  \newpage
\fi



%

\bibliographystyle{IEEEtrans} 
\bibliography{hoge} 

%




\end{document}